\documentclass[letter, twocolumn]{jpsj3} 
\usepackage{times}

\newcommand{\diff}{\mathrm{d}}
\newcommand{\imag}{\mathrm{Im}\,}

\newcommand{\imu}{\mathrm{i}}

\title{
Microscopic Mechanism for Staggered Scalar Order in PrFe$_4$P$_{12}$
}

\author{Shintaro \textsc{Hoshino}\thanks{E-mail: hoshino@cmpt.phys.tohoku.ac.jp}, Junya \textsc{Otsuki} and Yoshio \textsc{Kuramoto}}

\inst{Department of Physics, Tohoku University, Sendai 980-8578}

\recdate{\today}

\abst{
A microscopic model is proposed for the scalar order in 
PrFe$_4$P$_{12}$ 
where $f^2$ crystalline electric field (CEF) singlet and 
triplet states interact with two conduction bands. 
By combining dynamical mean-field theory and the continuous-time quantum Monte Carlo,
we obtain an electronic order with staggered Kondo and CEF singlets
with the total conduction number being unity per site.
The ground state becomes semimetallic provided that the two conduction bands have different occupation numbers.
This model naturally explains experimentally observed properties in the ordered phase of PrFe$_4$P$_{12}$ such as the scalar order parameter, temperature dependence of the resistivity, field-induced staggered moment, and inelastic features in neutron scattering.
The Kondo effect plays an essential role for ordering, in strong contrast with ordinary magnetic orders by the RKKY interaction.
}

\kword{
PrFe$_4$P$_{12}$,
Kondo effect,
crystalline electric field singlet,
dynamical mean-field theory
}

\begin{document}
\maketitle

Among Pr-based filled skutterudites that have attracted much recent attention, 
\cite{kuramoto09}
PrFe$_4$P$_{12}$ is one of the most interesting; it
shows a clear Kondo-like behavior in the resistivity, and undergoes a second-order phase
transition at $T_{\rm c}=6.5$K into non-magnetic staggered order.
In spite of the clear cusp in the magnetic susceptibility 
\cite{torikachvili87,aoki02} 
at $T_{\rm c}$ 
there is no magnetic moment in the ordered phase.
Instead, the neutron diffraction has revealed the field-induced staggered magnetic moment,\cite{iwasa08-2}
which shows the nonmagnetic character of the staggered order parameter.
Theoretical\cite{kiss05, sakai07} and experimental\cite{kikuchi07} studies have already identified the order parameter as a scalar, 
but the microscopic nature of this mysterious scalar order is yet to be clarified.

The inelastic neutron scattering experiments\cite{iwasa03, iwasa08, park08} on PrFe$_4$P$_{12}$ have probed
the broad quasi-elastic response above the transition temperature, which is characteristic of the Kondo effect.
Below $T_{\rm c}$, on the contrary, sharp inelastic excitations have been observed in addition to remaining quasi-elastic components.
The lineshape of the spectrum does not depend significantly on wave vector,\cite{iwasa_priv} but
the intensity of the scattering is stronger at the center of Brillouin zone than at the boundary.\cite{park08} 
The observed Kondo behaviors indicate a tendency toward itinerant $f$-electron state, while 
the inelastic feature may be interpreted as crystalline electric field (CEF) excitations.
Hence, PrFe$_4$P$_{12}$ can be regarded as a system located on the boundary between itinerant and localized characters of electrons.
Accordingly the ordered phase in PrFe$_4$P$_{12}$ may not be understandable in terms of the conventional RKKY interaction.

In the present paper, we propose microscopic mechanism for the scalar order in PrFe$_4$P$_{12}$.
In our previous work \cite{hoshino10}, we found a new electronic order with staggered Kondo and CEF singlets 
in a generalized Kondo lattice.
However, the previous paper obtained an insulating ground state, and did not
include comparison with experimental results in PrFe$_4$P$_{12}$.
In this paper, we demonstrate 
that the staggered Kondo-CEF singlet order is a plausible scenario for PrFe$_4$P$_{12}$ to 
explain its characteristic behaviors qualitatively. 
For example, 
we obtain semimetallic ordered state 
by taking a modified model for conduction electrons.
As a result, temperature dependence of the resistivity in PrFe$_4$P$_{12}$
is reasonably reproduced in the whole temperature range.
Furthermore, 
sharp inelastic peaks observed only below the transition temperature is naturally explained as local excitations in CEF sites generated by the ordering.

We take the singlet-triplet CEF states for low-lying $f^2$ states in PrFe$_4$P$_{12}$ as have been suggested in ref.\citen{otsuki05}.
For conduction bands, we adopt different orbitals $\gamma = 1,2$ 
since PrFe$_4$P$_{12}$ have 
two bands composed of $p$- and $d$-electrons.\cite{harima03} 
We take the following model:
\begin{align}
{\cal H} & = 
   \sum _{\mib{k}\gamma \sigma} ( \varepsilon _{\mib{k}\gamma} - \mu ) c_{\mib{k}\gamma\sigma} ^\dagger c_{\mib{k}\gamma\sigma}
 + J \sum _{i \gamma} \mib{S}_{\gamma i} \cdot \mib{s}_{{\rm c} \gamma i}    \nonumber \\
 & + 
\Delta  \sum_{i}\mib{S}_{1i} \cdot \mib{S}_{2i},
\label{eq_hamilt}
\end{align}
where $\mib{s}_{{\rm c}\gamma i}$ denotes the spin of conduction electrons at site $i$.
The local singlet-triplet states are represented by two pseudo spins $\mib{S}_1$ and $\mib{S}_2$, and the CEF splitting $\Delta$ is simulated by the coupling between them as shown in the third term.
The second term shows the Kondo exchange interaction between localized and conduction spins.

We use the dynamical mean-field theory (DMFT) formulated in the two-sublattice system in order to discuss the staggered order.
In this theory, the lattice is mapped to the effective impurity problem with full on-site correlations.\cite{georges96}
The continuous-time quantum Monte Carlo method (CT-QMC)\cite{rubtsov05, werner06, otsuki07} is applied as the impurity solver.
For analytic continuation from imaginary axis onto real one, we employ the Pad\'{e} approximation.

\begin{figure}
\begin{center}
\includegraphics[width=75mm]{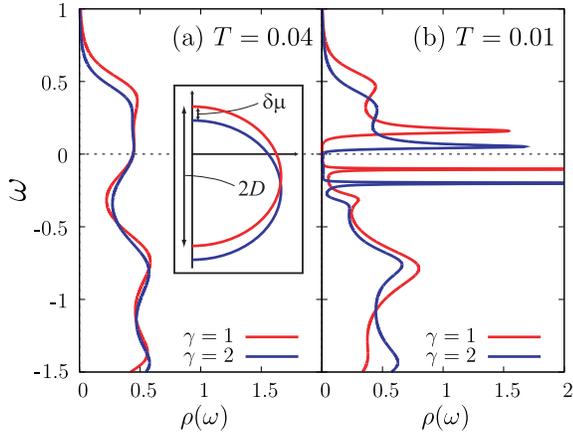}
\caption{
(color online)
Density of states in (a) disordered phase at $T=0.04$ and (b) ordered phase at $T=0.01$.
The inset in (a) illustrates the bare DOS of conduction bands with different scales.
}
\label{fig_dos}
\end{center}
\end{figure}

For the spectrum of each conduction band, we take 
$\varepsilon_{\mib{k}1} = \varepsilon_{\mib{k}} + \delta \mu /2$
and
$\varepsilon_{\mib{k}2} = \varepsilon_{\mib{k}} - \delta \mu /2$
with $\delta \mu$ controlling the asymmetry.
Here $\varepsilon_{\mib{k}}$ gives the semi-circular density of states (DOS) defined by
\begin{align}
\rho_0 (\varepsilon) = \frac{2}{\pi D^2} \sqrt{ D^2 - \varepsilon ^2 }
. \label{eq_bethe_dos}
\end{align}
The band width is given by $2D$ with $D=1$ as a unit of energy.
Throughout this paper, the number of conduction {\it holes} per site is fixed at $n_{\rm c} = n_{\rm 1} + n_{\rm 2} = 1$, which corresponds to the $3/4$ filling of two conduction bands.
In this paper, we take the parameters such 
as $J=0.8$, $\Delta = 0.2$ and $\delta\mu =0.1$.
Without interaction, we obtain $n_{\rm 1} \simeq 0.56 > n_{\rm 2} \simeq 0.44$ at zero temperature.
Note that the ratio $n_1/n_2$ depends on temperature.
The DOS without interaction is schematically shown in the inset of Fig. \ref{fig_dos}.
These bands simulate the unit number of conduction holes 
in LaFe$_4$P$_{12}$\cite{sugawara00}, and make the starting point for 
PrFe$_4$P$_{12}$ with localized picture of $f$ electrons.
The Kondo temperature for the impurity system is given by $T_{\rm K} \equiv D \exp (-2D/J) \simeq 0.082$, which is comparable to $\Delta$.
The large value of $J$ is taken simply for easier numerical calculation.
Because of strong renormalization effect, however, the characteristic temperatures are still much smaller than the band width $2D$.
Hence, the low-energy behaviors are not sensitive to the choice of the bare parameters.

Under the present condition, the system undergoes the staggered Kondo-CEF singlet ordered phase below $T = T_{\rm c} \sim 0.031$ by the second-order transition.
In this phase, the conduction holes accumulate more on the Kondo-singlet site than on the CEF-singlet site.\cite{hoshino10}
Clearly the symmetry of the order parameter is a scalar.
The origin of this ordering can be understood from the strong coupling limit with respect to $J$.\cite{hirsch84, sigrist92, otsuki09}
In this limit, each site forms the Kondo singlet in presence of one conduction electron, or CEF singlet with fully-occupied conduction state.
The second-order perturbation with respect to the hopping of conduction electrons gives the effective repulsive interaction between Kondo singlets, which favors the staggered order with Kondo and CEF singlets.
This order is induced by the Kondo effect, and the RKKY interaction is irrelevant here.
As is clear from the formation of the CEF singlet, the present order is characteristic of the systems with $f^2$ configuration.
In terms of itinerant and localized characters, the staggered Kondo-CEF singlet order can be regarded as alternating sites with itinerant and localized $f$ electrons.
We note that the present order 
has nothing to do with nesting property of the half-filled single conduction band,
which can be relevant in other skutterudites such as PrRu$_4$P$_{12}$.\cite{harima03}
Instead, it is essential that we have two conduction bands each of which is $3/4$-filled on the average.

Now we discuss the physical properties of the order and the possible relevance to the experiments in PrFe$_4$P$_{12}$.
Figure \ref{fig_dos} shows the DOS
both in (a) the disordered phase above $T_{\rm c}$ 
and (b) the ordered phase.
For $T>T_{\rm c}$ as shown in 
Fig.\ref{fig_dos}(a), the DOS at the Fermi level is finite.
The shape of DOS is modified much from the bare one shown in the inset because of the strong  interaction effect.
At $T=0.04$, the difference of numbers of holes between two bands is given by $n_1 - n_2 \simeq 0.094$.
Below the transition temperature as shown in Fig.\ref{fig_dos}(b), 
on the other hand, 
a pseudo-gap structure arises near the Fermi level with a sharp peak at each edge.
The peaks above the Fermi level come from Kondo-singlet sites, and the ones below are due to the formation of CEF-singlet sites.
In the ordered phase, the numbers of conduction holes come closer to each other 
with $n_1 - n_2 \simeq 0.004$ at $T=0.01$.
The DOS at the Fermi level 
is tiny but finite, and hence the ground state becomes metallic.

Next we derive the electrical resistivity $R$ from the current-current correlation function.
According to the result for the Hubbard model in infinite dimensions,\cite{pruschke93} 
we take the following form for the current operator ${\cal J}$ for the two-sublattice system:
\begin{eqnarray}
 {\cal J}  = {\sum_{\mib{k}}} ' \sum_{\gamma \sigma} v_{\mib{k}\gamma} (
c_{\mib{k}{\rm A} \gamma \sigma}^\dagger c_{\mib{k} {\rm B} \gamma \sigma} + {\rm h.c.}
),
\end{eqnarray}
where $c_{\mib{k}\lambda \gamma \sigma}$ is the annihilation operator of conduction electrons at sublattice $\lambda = {\rm A, B}$ with spin $\sigma$ and orbital $\gamma$.
The form of $ {\cal J} $ derives from the nearest-neighbor hopping term of conduction electrons in the bipartite lattice, and is off-diagonal with respect to sublattice index $\lambda$.
The summation $\sum'_{\mib{k}}$ is taken over the reduced Brillouin zone of the superlattice.
Then we calculate the quantities 
\begin{align}
\Pi   (\imu \nu_m) &= 
\int _0^\beta \diff \tau \langle T_\tau  {\cal J} (\tau)  {\cal J} \rangle \exp(\imu \nu _m \tau) 
\\
R^{-1} &= A \lim_{\omega\rightarrow 0} 
\imag \Pi  (\omega + \imu \delta)/\omega
\label{eqn_corr} 
\end{align}
where $\nu _m = 2m\pi T$ is a bosonic Matsubara frequency and $A$ is a constant.
In the framework of the DMFT, the vertex correction can be neglected,\cite{khurana90}
and we evaluate the simple particle-hole bubble in eq.(\ref{eqn_corr}).
Assuming that the relevant energy range is much smaller than the band width,
we replace the velocity $v_{\mib{k}\gamma}$ by a constant.

\begin{figure}
\begin{center}
\includegraphics[width=75mm]{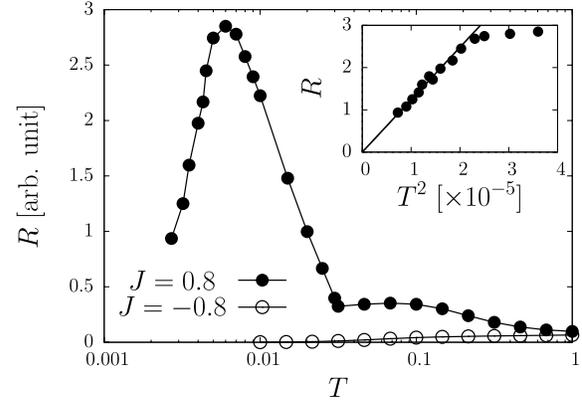}
\caption{
Temperature dependences of the resistivity $R$ both above and below the transition temperature $T_{\rm c} \sim 0.031$.  The inset shows the low temperature behavior.
For comparison, the ferromagnetic case with $J=-0.8$ is also shown.
}
\label{fig_cond}
\end{center}
\end{figure}
Figure \ref{fig_cond} shows the electrical resistivity $R$.
We also show the result with $J=-0.8$ for comparison, where the Kondo effect does not operate.
At high temperatures,
the resistivities with $J=0.8$ and $-0.8$ tend to the common value obtained by the Born approximation.
The resistivity for $J=0.8$ increases with decreasing $T$, while it is not the case for $J=-0.8$.
Hence the increase in the resistivity at $T>T_{\rm c}$ is indeed due to the Kondo effect.
Below the transition temperature $T_{\rm c}$, on the other hand, 
the resistivity shows sharp increase, 
and then decreases after showing the maximum around
$T\sim 0.19 T_{\rm c}$.
The sharp increase reflects the formation of energy gap at the Fermi level,
while the decrease shows
the metallic behavior at sufficiently low temperatures.
The inset in Fig. \ref{fig_cond} magnifies the low temperature regime with $T\lesssim 0.006$, where the Fermi liquid behavior $R \propto T^2$ is found.
Experimentally, the resistivity in PrFe$_4$P$_{12}$  increases more abruptly\cite{torikachvili87, sato00} just below $T_{\rm c}$ than shown in Fig. \ref{fig_cond}.
This may be due to the short-range fluctuation of Kondo and CEF sites,
which causes the disorder scattering, but is not included in our two-sublattice DMFT.

We now discuss magnetic response of the system.
In filled skutterudites with the CEF symmetry $T_h$, the triplet states consist of linear combinations of two triplets under the cubic symmetry:
$\Gamma_4$ and $\Gamma_5$ states.
With use of the mixing parameter $d$ between them, we write the first-excited triplet states as
$|{\rm t}\rangle = \sqrt{1-d^2} | \Gamma_5  \rangle + d | \Gamma_4  \rangle$.
The dipole operator is represented as $\mib{M} = a_1 \mib{S}_1 + a_2 \mib{S}_2$ with $a_1, a_2$ being a function\cite{shiina04} of $d$.
The Curie constant is given by $C = (a_1^2 + a_2^2)/4$.
According to ref. \citen{kuramoto05}, the triplet in PrFe$_4$P$_{12}$ is mainly composed of $\Gamma_4$.
Hence we choose $d=0.8$ in this paper, which gives\cite{shiina04} $a_1 = 3.29$ and $a_2 = -0.85$.

Let us define
{\it partial}
magnetic susceptibilities by
\begin{align}
\chi_M^{\lambda\lambda'} (\mib{q}, \imu \nu_m) = \int _{0} ^{\beta} \diff \tau \langle T_\tau M_z^\lambda (\mib{q},\tau) M_z^{\lambda'} (-\mib{q}) \rangle 
 \exp (\imu \nu_m \tau )
\label{eq_mag} ,
\end{align}
where $M_z^\lambda$ denotes the magnetic moment at sublattice $\lambda$.
The wave vector $\mib{q}$ in eq. (\ref{eq_mag}) belongs to the reduced Brillouin zone.
The uniform and staggered susceptibilities
in the original Brillouin zone
are then given by
\begin{align}
&\chi_M (\mib{0}, \imu \nu_m) =  \frac{1}{2} \sum_{\lambda\lambda'}\chi_M^{\lambda\lambda'} (\mib{0}, \imu \nu_m)
, \label{eq_unif} \\
&\chi_M (\mib{Q}, \imu \nu_m) =  \frac{1}{2} \sum_{\lambda\lambda'}{\rm sgn_{\lambda}}{\rm sgn_{\lambda'}}\,
\chi_M^{\lambda\lambda'} (\mib{0}, \imu \nu_m) 
, \label{eq_stag}
\end{align}
with sgn$_{\rm A} = 1$ and sgn$_{\rm B} = -1$.
We also introduce the local susceptibility 
$\chi_M^{{\rm loc}, \lambda}$ for sublattice $\lambda$, 
and their average by
\begin{align}
\chi_M^{\rm loc} (\imu \nu_m ) & \equiv 
\frac{1}{N} {\sum_{\mib{q}}}' \sum_{\lambda}
\chi_M^{\lambda\lambda} (\mib{q}, \imu \nu_m )
\nonumber \\
 & = \frac{1}{2} \left[ 
 \chi_M^{{\rm loc, A}}(\imu \nu_m ) + \chi_M^{{\rm loc, B}}(\imu \nu_m ) \right]
, \label{eqn_loc_sus}
\end{align}
where $N$ is the number of sites.
The susceptibilities (\ref{eq_unif}), (\ref{eq_stag}) and (\ref{eqn_loc_sus}) are calculated from local two-particle correlation functions in the two-sublattice system.\cite{hoshino10, otsuki09-2}

\begin{figure}
\begin{center}
\includegraphics[width=80mm]{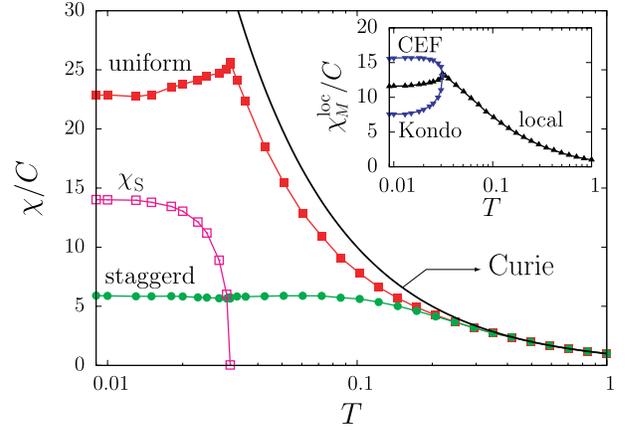}
\caption{
(color online)
Temperature dependences of static magnetic response: 
Curie law, uniform, and staggered susceptibilities. The field-induced staggered susceptibility $\chi_{\rm S}$ is defined in the text.
The inset shows
local susceptibilities for Kondo- and CEF-singlet sites, together with their average.
}
\label{fig_mag_sus}
\end{center}
\end{figure}

Figure \ref{fig_mag_sus} shows the numerical results for static susceptibilities.
The uniform susceptibility $
\chi_M (\mib{0}, 0) $
shows the clear cusp at $T_{\rm c}$,
as in antiferromagnetic transition.
Note that the
experimentally
observed magnetic susceptibility also shows a cusp \cite{torikachvili87,aoki02}.
On the contrary,
the staggered susceptibility $
\chi_M (\mib{Q}, 0)$ does not show any distinct anomaly at the transition temperature.
The magnitude of $\chi_M (\mib{Q}, 0)$
is much smaller than the uniform susceptibility.
Namely, the staggered scalar order has nothing to do with the antiferromagnetic correlation.
The inset shows the local susceptibility that
splits into two different values between Kondo- and CEF-singlet sites.
The CEF-singlet site has
larger susceptibility than the Kondo-singlet site.

We can also discuss
the field-induced staggered moment observed in the neutron diffraction.\cite{iwasa08}
The relevant susceptibility is the 
staggered magnetic response against uniform magnetic field.  Namely, we introduce
\begin{align}
\chi_{\rm S} 
 = \frac{1}{2} \sum_\lambda {\rm sgn_{\lambda}} \, \chi^{\lambda \lambda}_M (\mib{0},0)
,
\end{align}
which vanishes in the disordered phase.
Here we have used the relation $\chi_M^{\rm AB} = \chi_M^{\rm BA}$.
As shown in Fig. \ref{fig_mag_sus}, we obtain finite $\chi_{\rm S}$
in the ordered phase, which means that  the staggered magnetic moment appears under uniform magnetic field.
The magnitude of the moment is 61\%
of the uniform moment for $T=0.01$ under the present condition.

\begin{figure}
\begin{center}
\includegraphics[width=80mm]{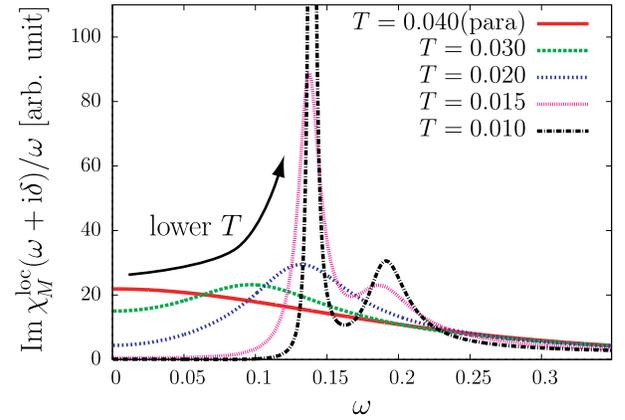}
\caption{
(color online)
Local magnetic spectrum at several temperatures.
}
\label{fig_mag_spect}
\end{center}
\end{figure}

The temperature dependence of the local magnetic spectrum $\imag \chi ^{\rm loc}_M (\omega + \imu \delta) / \omega$ is shown in Fig. \ref{fig_mag_spect}.
Above $T_{\rm c}$, the spectrum shows broad quasi-elastic peak which is characteristic of the Kondo effect, even though the CEF splitting is present.
In the ordered phase, on the contrary, the sharp inelastic peak at $\omega \sim 0.14$ grows with decreasing temperature.
This peak is due to  emergence of the CEF-singlet site, and the corresponding energy is smaller than the magnitude of the original CEF splitting, owing to the interaction effect.
The other and wider inelastic peak at $\omega \sim 0.19$ originates from the Kondo-singlet site where
the DOS has the gap structure.
These results are consistent with neutron scattering results\cite{iwasa03}
in that the broad quasi-elastic peak is observed in the disordered phase, and sharp inelastic peaks appear in the ordered phase. 

The neutron scattering on single crystal\cite{iwasa_priv} shows that the inelastic peak has almost no dispersion in the ordered phase.
On the other hand, the intensity of the spectrum depends on the wave vector; 
the intensity is the largest at the zone center and is the weakest at the zone boundary.
Note that the sum rule of the inelastic spectrum with fixed wave number gives the static susceptibility.
As dynamical magnetic response, we have 
derived only the local spectrum in this paper.
However,
the energy-integrated neutron spectrum
can be compared with 
the wave-vector dependent static magnetic susceptibility. 
From the results shown in Fig.\ref{fig_mag_sus},
it is clear that the dependence 
on wave vector of the neutron intensity is consistent with that of the static susceptibilities.

Let us comment on 
other interesting aspect of PrFe$_4$P$_{12}$ in the light of the present results.
Since the formation of the Kondo singlet between conduction and $f$ electrons is essential, the order should be fragile against disorder.
In fact, slight substitution of Pr by La easily destroys the order as observed in Pr$_{1-x}$La$_x$Fe$_4$P$_{12}$.\cite{tayama07}
On the other hand, it has been reported that antiferromagnetic insulator is realized under high pressure.\cite{hidaka05, osakabe10}
This insulating behavior implies the nesting properties of conduction bands as in PrRu$_4$P$_{12}$.
\cite{harima03}
We ascribe this behavior to the change of the conduction bands.  Namely,
if one of the conduction bands disappears under pressure, 
the remaining band is half-filled, and strong nesting property may emerge.
Then it is not surprising that we have the insulating ordered phase with antiferromagnetism.

Some experiments show the need of including higher CEF levels in PrFe$_4$P$_{12}$.
Under high magnetic field along $(1,1,1)$ direction, new ordered phases and non-Fermi liquid behavior have been found.\cite{tayama04}
The special property with the $(1,1,1)$ field cannot be understood in the singlet-triplet CEF levels, but 
the full CEF model reproduces the level crossing only along $(1,1,1)$ field \cite{kiss05}. 
Some refinement of the model such as the quasi-sextet CEF model including higher doublet 
will improve the correspondence between theory and experiments.

In conclusion, we have demonstrated the relevance of a variant of the Kondo lattice model to PrFe$_4$P$_{12}$ using the DMFT+CT-QMC.
Our scenario of the staggered Kondo-CEF singlet order can naturally explain many experimental results in PrFe$_4$P$_{12}$ such as the scalar order parameter, temperature dependence of the resistivity, and sharp inelastic magnetic response observed only in the ordered phase.

The authors are grateful to K. Iwasa, A. Kiss and H. Hidaka for fruitful discussions.
One of the authors (S. H.) is supported by the global COE program of MEXT, Japan.
This work was partly supported by a Grant-in-Aid for Scientific Research on Innovative Areas ``Heavy Electrons" (No 20102008) of MEXT.

\label{lastpage}
\clearpage

\end{document}